\documentclass[a4paper,10pt,english]{article}
\usepackage{epstopdf}
\usepackage{subfigure}
\usepackage{latexsym}
\usepackage{natbib}
\usepackage{bbm}
\usepackage{amsmath,amssymb,amsfonts}
\usepackage{graphicx}
\usepackage{wrapfig}
\usepackage{amsmath}
\usepackage{graphics}
\usepackage[latin1]{inputenc}
\usepackage[a4paper]{geometry}
\begin{document}



\title{ \Large Generalized Dynkin game of switching type representation for
defaultable claims in presence of contingent CSA }

\maketitle


\begin{center}
  GIOVANNI MOTTOLA\footnote{Mail: g.mottola@be-tse.it}\\
  Sapienza University of Rome\\
  Piazzale Aldo Moro\\
  00185, Rome, Italy
\end{center}


\noindent
\begin{abstract}
\noindent We study the solution's existence for a generalized Dynkin game of switching
type which is shown to be the natural representation for general defaultable OTC contract with contingent CSA. This is a theoretical counterparty risk
mitigation mechanism that allows the counterparty of a general OTC contract to switch from zero
to full/perfect collateralization and switch back whenever she wants until contract maturity paying
some switching costs and taking into account the running costs that emerge over time. In this
paper we allow for the strategic interaction between the counterparties of the underlying contract,
which makes the problem solution much more tough. We are motivated in this research by the
importance to show the economic sense - in terms of optimal contract design - of a contingent
counterparty risk mitigation mechanism like our one. In particular, we show that the existence
of the solution and the game Nash equilibrium is connected with the solution of a system of
non-linear reflected BSDE which remains an open problem. We then provide the basic ideas to numerically search the game equilibrium via an \emph{iterative optimal stopping} approach and we show the existence of the solution for our problem under strong condition, in the so called \emph{symmetric case}.



\end{abstract}

\section{\Large Introduction}
\subsection{\large Aim of the work }
In this paper we analyze a theoretical contract in which counterparties want to set a contingent CSA (\emph{credit support annex})  in order to gain  the flexibility and the possibility to manage optimally the counterparty risk. 
We refer specifically to a contingent risk mitigation mechanism that allows the counterparties to switch from zero to full/perfect collateralization (or even partial) and switch back whenever until maturity $T$ paying some \emph{switching costs }and taking into account the\emph{ running costs } that emerge over time.
The running costs that we model and consider in the analysis of this problem are - by one side - those related to CVA namely\emph{ counterparty risk hedging costs} and - by the other side - the \emph{collateral and funding/liquidity costs} that emerge when collateralization is active.\\
We can summarize the characteristics and the basic idea underlying the problem  - that we show to admit a natural formulation as a \emph{stochastic differential game of switching type} -   through the so defined \emph{contingent CSA scheme} shown below (Fig. 1.1), in which - by considering also the funding issue in the picture - is present a third party, an \emph{external funder} assumed \emph{default free} ($\lambda =0$) in order to reduce dimension and technical issues.\\
\begin{figure}[h!]
  \centering
  \includegraphics[  scale=0.5]{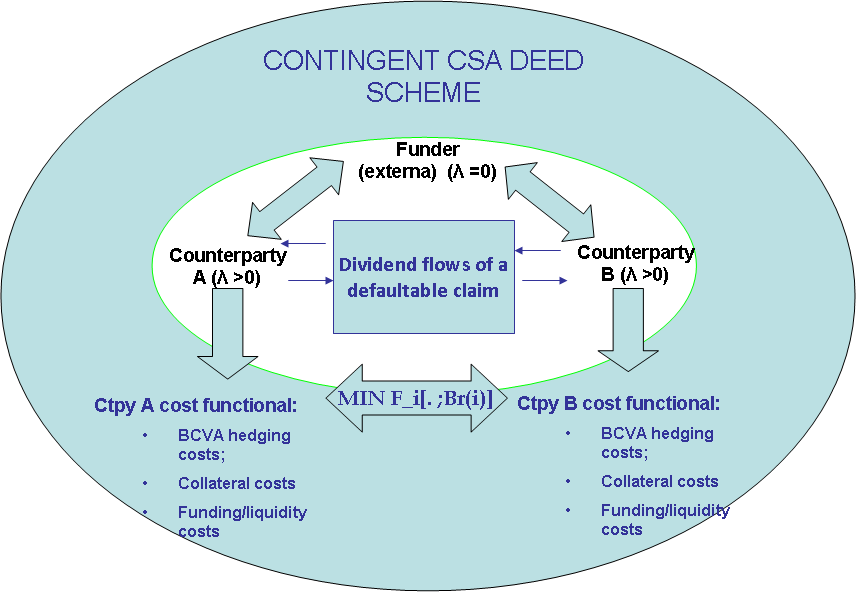}
\end{figure}
Here, motivated by the results obtained in the unilateral case\footnote{We refer to the stochastic optimal control formulation highlighted in the third section of Mottola (2013)}, we analyze the problem in a generalized setting allowing for the strategic interplay between the parties of the contingent CSA scheme. \\
This has lead us to study  the solution's existence and the equilibrium for the \emph{stochastic differential game of switching type} that we are going to define in \emph{section two}. We can anticipate that our game solution (its existence and uniqueness) remains an open problem, that deserves further studies and research.
We also highlight the basic ideas of the game's numerical solution  via  iterative optimal stopping
approach. Then further analysis on game equilibrium  are taken over in \emph{section three } and some model applications in \emph{section four}.
\subsection{\large Literature review}


The body of literature regarding stochastic differential games theory is very wide. The roots of the theory are founded in the pioneering works of \cite{Jvn44} - for (mainly cooperative) \emph{ zero sum game} - and \cite{Nas50} - for (non cooperative) \emph{non-zero sum game} -  and the work of Isaacs (see \cite{Isa99}) who studied for first the \emph{differential games} in a deterministic setting. In the stochastic framework it is worth of mention the seminal work of Eugene Dynkin (\cite{Dyn67})  who firstly analyzed stochastic differential games where the agent control set is given by \emph{stopping times}, so that these ''games on stopping'' are known as \emph{Dynkin games} in his honor.\\
Certainly, it is impossible to mention here all the numerous important contribution in this field of research and to give a systematic account of the theory and of the literature. For a complete treatment of the different type of games we refer to the book of Isaacs on differential games. So we give in the following just a simplified classification restricting ourselves to the literature more related to our stochastic switching control problem.\\ This classification is based on the following main categories:\\

\indent a)\emph{game and equilibrium type}: it includes zero and non-zero-sum  games whose solution can be searched mainly in terms of cooperative or non-cooperative (Nash) equilibrium. This depends also on the characteristics of the game which are mainly the system dynamic - that can be \emph{markovian/non markovian} - and controls which can be state controls, stopping controls (as in Dynkin games) or both which are called \emph{mixed control and stopping}. Also the number of player is relevant, here we focus on the case $p=2$.\\

\indent  b) \emph{solution approaches}: these are mainly the \emph{analytical approach}  that allows - typically under a markovian framework - to formulate the stochastic differential game (SDG) as a system of (second order) \emph{Hamilton-Jacobi-Bellman equations} or \emph{variational inequalities} to solve, proving existence and (possibly) uniqueness of the solution, namely of the equilibrium of the game. The main solution techniques are the ones related to PDE theory, the dynamic programming principle  and viscosity solution. \\
      Worth of mention, from the analytical point of view, are the important works of \cite{bens77} that for first showed the existence of a Nash equilibrium for a non-zero-sum SDG with stopping $\{\tau_{1},\tau_{2}\}$ as controls, formulating the problem as a system of quasi-variational inequalities (solved through \emph{fixed point} methods), assuming continuous and bounded running rewards and terminal rewards; instead \cite{Fle89} for first showed the existence and uniqueness of the solution/equilibrium for zero-sum SDG through dynamic programming and viscosity solution approach. These techniques have become very popular and used in the recent literature given the deep connection with  probabilistic tools, as we have already mentioned in chapter five.\\
      In fact, the \emph{probabilistic approach} is the other more general one\footnote{In fact, it allows to deal also with general non markovian dynamics for the system state variables. } that makes use of the martingale (also via \emph{duality methods}) and  Snell envelope theory  and, in addition, of the deep results of the forward-backward SDE theory in order to derive existence and uniqueness of the optimal control/stopping strategy for the game.\\
      The main works worth of mention - other that the already mentioned work of \cite{Civ96} that for first highlights the connection between the solution of zero-sum Dynkin game and that of \emph{doubly reflected BSDE} (other than its analytical solution) - are that of:  \cite{Ham98} that shows how the solution of a SDG is related to that of a backward-forward SDE; \cite{Ham00} that extend the analysis through reflected BSDE to ''\emph{mixed game}'' problems; \cite{Elk03} that generalize the existence and uniqueness results for zero and non-zero-sum game with ''\emph{risk sensitive}'' controls; \cite{Ham08} that use Snell envelope technique to show the existence of a Nash equilibrium for non-zero-sum Dynkin game in a non-markovian framework and \cite{Ham10}  that tackle the solution of a general switching control problem via systems of interconnected (nonlinear) RBSDE (with \emph{oblique reflection}).\\

      To conclude the section we recall also the monographic work of \cite{Pha09} and \cite{Yon99} for a clear and detailed analysis of backward SDE and we remark also that much of these literature and works have been inspired by financial valuation problems. We refer mainly to the \emph{american game option} problem as defined in \cite{Kif00} (also known as \emph{israeli option}). This has given impulse to the literature  related mainly to \emph{convertible and switchable bond valuation }whose solution can be related to that of a zero-sum Dynkin game.

\subsection{ \large Some examples od Dynkin games}
Let us briefly remind that stochastic differential games are a family of dynamic, continuous
time versions of \emph{differential games} (as defined by Isaacs) incorporating randomness in both the
states and the rewards. The random states are described typically by adapted diffusion processes
whose dynamics are known (or assumed). To play a game, a player receives a \emph{running reward}
cumulated at some rate till the end of the game and a \emph{terminal reward} granted at the end of
the game. The rewards are related to both the \emph{state process} and the \emph{controls} at the choice of
the players, as deterministic or random functions or functionals of them. A control represents
a player action in attempt to influence his rewards. Assuming his rationality, a player acts in
the most profitable way based on his knowledge represented by his information filtration. Before
starting the formulation and the analysis of our generalized \emph{Dynkin game of switching type}, let us
recall - in a markovian framework - some example for both a non-zero-sum and zero-sum Dynkin
game (with $p = 2$) and the relative equilibrium characterization. For all the details we refer in particular to the works of \cite{bens77} and \cite{Fle89}\\

-\emph{Non-zero-sum Dynkin game}: Given a standard probability space represented by the triple $(\Omega,\mathcal{F},\mathbb{P}) $ where we define  $W = (W_{t})_{0\leq t\leq T}$ be a standard $d$-dimensional Brownian motion adapted to the space filtration. Assuming as true the usual conditions on the drift function $\mu(.)$ and volatility function $\sigma(.)$, such that the following SDE admits a unique solution
      \begin{eqnarray*}
        dy(t) &=& \mu(y(t),t)dt+\sigma(y(t),t)dW(t), \;\; t\in[0,T]\\
        y(0) &=& y_{0}.
      \end{eqnarray*}
  Let (for $p = 1,2$) $f_{p}(y,t)$  the running reward function and $\phi_{p}(y,t)$, $\psi_{p}(y,t)$ the reward function obtained by the players upon stopping the game be \emph{continuous} and \emph{bounded} in $\mathbb{R}^{d}\times[0,T]$, with $f_{p}\in\mathbb{L}^{2}$ square integrable and $\psi_{p}\leq\phi_{p}$ for all $(y,t)\in \mathbb{R}^{d}\times[0,T]$. Then let $g_{p}(y(T))$  the  terminal reward function also continuous and bounded. \\
  In a game of this kind, the two players have to decide optimally when to stop the game finding the optimal control given  by the stopping times $(\tau_{1},\tau_{2})$ that give the maximum expected reward. So let us set the payoff functional for the two players of this Dynkin game as follows
  \begin{eqnarray*}
    J^{p}(y,\tau_{1},\tau_{2}) &=&\mathbb{E}\bigg[ \int_{t}^{\tau_{1}\wedge\tau_{2}\wedge T} f_{p}(y(s),s)ds+ \mathbbm{1}_{\{\tau_{i}<\tau_{j} \}} \phi_{p}(y(\tau_{i}), \tau_{i}) \\
     &+ & \mathbbm{1}_{\{\tau_{i} \geq\tau_{j}, T>\tau_{j}  \}} \psi_{p}(y(\tau_{j}), \tau_{j}) +  \mathbbm{1}_{\{\tau_{1} =\tau_{2}= T  \}} g_{p}(y(T)) \bigg] \:\; for\; j\neq i(\:\in\{1,2\}),
  \end{eqnarray*}
 and for $(t\leq \tau_{i}\leq T)$. Being in a non-zero-sum game with the players that aims to maximize their payoff $J^{p}(.)$ without cooperation, the problem here is to find a \emph{Nash equilibrium point} (NEP) for the game, that is to determine the couple of optimal stopping times $(\tau_{1}^{*},\tau_{2}^{*})$ such that
 \begin{eqnarray*}
   J^{1}(y,\tau_{1}^{*},\tau_{2}^{*}) &\geq& J^{2}(y,\tau_{1}^{},\tau_{2}^{*}), \;\: \forall \: \tau_{1},\tau_{2} \in [t,T] \\
   J^{2}(y,\tau_{1}^{*},\tau_{2}^{*}) &\geq& J^{1}(y,\tau_{1}^{*},\tau_{2}^{}), \;\: \forall \:\tau_{1},\tau_{2}\in [t,T]
 \end{eqnarray*}
namely the supremum of the payoff functional over the stopping time set. In other words, the NEP implies that every player has no incentive to change his strategy given that the other one has already defined optimally his strategy.\\
This type of game, as shown in \cite{bens77}, has an analytical representation given by a system of \emph{variational inequalities} but it admits also a stochastic counterpart through \emph{system of BSDE with reflecting barrier}. We return to its formal definition in the next section in relation to our problem.
The Nash equilibrium defined above can be fairly generalized in the case of \emph{mixed game of control and stopping}. We show this below in relation to zero-sum games.\\

- \emph{Zero-sum mixed game}: A zero-sum game is characterized by the antagonistic interaction of the players that in this case has the same payoff functional but their objective are different because for one player the payoff is a reward (let's think typically at the buyer of a convertible bond) that he wants to \emph{maximize}, while for the other one is a cost that he intends to \emph{minimize}. \\
    In the generalized case of mixed games of control and stopping, the set of control will be enriched by the $\mathcal{F}_{t}$-progressively measurable process $(\alpha_{t})_{t\leq T}$ and $(\beta_{t})_{t\leq T}$ that are the intervention function namely the state controls respectively for the player $p_{1}$ and $p_{2}$. In addition, the players have to decide optimally when to stop the game  setting the stopping times $\tau$ (for $p_{1}$) and $\sigma$ (for $p_{2}$). Indeed, the system dynamic being controlled by the agents can be expressed as the following \emph{controlled diffusion} (remaining in a markovian framework):
      \begin{eqnarray*}
        dy(t)^{\alpha,\beta} &=& \mu(t,y_{t}^{\alpha,\beta},\alpha_{t},\beta_{t})dt+\eta(t,y_{t}^{\alpha,\beta},\alpha_{t},\beta_{t})dW(t), \;\; t\in[0,T]\\
        y(0)^{\alpha,\beta} &=& y_{0}.
      \end{eqnarray*}
The zero-sum game payoff being the same for both the players will be
  \begin{eqnarray*}
    \Gamma(\alpha,\tau; \beta,\sigma) &:=& \mathbb{E}\bigg[ \int_{t}^{T\wedge\tau\wedge\sigma} f(s,y_{s}^{\alpha,\beta},\alpha_{s},\beta_{s})ds + \mathbbm{1}_{\{\tau\leq\sigma<T \}} \phi(\tau,y_{\tau}^{\alpha,\beta}) \\
     &+ & \mathbbm{1}_{\{\sigma <\tau  \}} \psi( \sigma,y_{\sigma}^{\alpha,\beta}) +  \mathbbm{1}_{\{\tau =\sigma= T  \}} g(y^{\alpha,\sigma}(T)) \bigg] \:\;  (t\leq \tau\leq \sigma <T)
  \end{eqnarray*}
  where the running and reward functions are intended to be the same as in the non-zero-sum case but clearly now they are the same for both players.\\
 The solution of this SDG is typically tackled by studying the upper and lower \emph{value function}  of the players, which are
 \begin{eqnarray*}
    \mathcal{U}(t,y)&:=& \sup_{\alpha}\inf_{\beta} \sup_{\tau}\inf_{\sigma} \Gamma(\alpha,\tau; \beta,\sigma)\;\;(upper \:value\:p1) \\
    \mathcal{L}(t,y)&:=& \inf_{\beta}\sup_{\alpha} \inf_{\sigma}\sup_{\tau} \Gamma(\alpha,\tau; \beta,\sigma)\;\;(lower \:value\:p2)
 \end{eqnarray*}
    Under some standard condition on the reward function and on controls, the problem has been tackled analytically representing the lower and upper value of the game as a system of nonlinear PDE with two obstacles/barriers, defined as follows
    \[
  \begin{cases}
       \min \bigg\{ u(t,y) -  \phi(t,y), \max\bigg\{ \frac{\partial u}{\partial t}(t,y) - H^{-}(t,y,u,Du,D^{2}u), u(t,y) - \psi(t,y)  \bigg\}\bigg\}=0    \\
       u(T,y) = g(y),
        \end{cases}\\
\]
\[
  \begin{cases}
       \min \bigg\{ v(t,y) -  \phi(t,y), \max\bigg\{ \frac{\partial v}{\partial t}(t,y) - H^{+}(t,y,v,Dv,D^{2}v), v(t,y) - \psi(t,y)  \bigg\}\bigg\}=0\\
       v(T,y) = g(y),
        \end{cases}\\
\]\\
where $H^{+}(.)$ and $H^{-}(.)$ are the \emph{hamiltonian operators} (as defined in chapter four) associated to the upper and lower value function of the SDG. To solve the system,  the unknown solution function $u$ and $v$ can be shown (under some techinical assumptions) to be \emph{viscosity solutions}  of the above two PDE with obstacles and to coincide with the value functions $\mathcal{U}(t,y)$ and $\mathcal{L}(t,y)$ of the game.\\
In particular, when the \emph{Isaacs condition} holds, namely 
\begin{equation*}
    H^{-}(t,y,u,q,X)=H^{+}(t,y,u,q,X)
\end{equation*}
then the two solutions coincide and the SDG has a value namely
\begin{equation*}
    V := \sup_{\alpha}\inf_{\beta} \sup_{\tau}\inf_{\sigma} \Gamma(\alpha,\tau; \beta,\sigma) =  \inf_{\beta}\sup_{\alpha} \inf_{\sigma}\sup_{\tau} \Gamma(\alpha,\tau; \beta,\sigma)
\end{equation*}
which is called the \emph{saddle point equilibrium} of the mixed zero-sum game. \\
We mention also that in this case, as in the non-zero-sum case, the SDG has a stochastic representation which is expressed in terms of a \emph{doubly reflected BSDE} ($2RBSDE$)\footnote{Its connection with the analytical representation and viscosity solution of PDE with obstacles, as already mentioned, has been established in the work of \cite{Civ96}. }. In particular, setting the terminal
reward $\xi$, the \emph{early exercise rewards} $\phi_{t} = U_{t}$ and  $\psi_{t} = L_{t}$ which represent the two barriers of the value process, therefore the \emph{generator function} $f$, the \emph{value process} $Y_{t}$, $Z_{t}$ which is \emph{the conditional expectation/volatility process} that helps $Y_{t}$ to be $\mathcal{F}_{t}$-measurable and $K$ the \emph{compensator process}, it can be shown that the following $2RSBDE$\footnote{This is established in the work of \cite{Civ96}.} solution
\[
  \begin{cases}
       Y_{t} = \xi + \int_{t}^{T}f(s,Y_{s},Z_{s})+ (K_{T}^{+}-K_{t}^{+}) - (K_{T}^{-}-K_{t}^{-})- \int_{t}^{T}Z_{s}dW_{s}\;\;\forall t\leq T\\
       L_{t}\leq Y_{t}\leq U_{t},\; \forall t\leq T,\;\: \\
       \int_{0}^{T}(Y_{s}-L_{s})dK_{s}^{+}= \int_{0}^{T}(U_{s}-Y_{s})dK_{s}^{-}=0
        \end{cases}\\
\]\\

\section{ \Large Defaultable Dynkin game of switching type}

\subsection{\large Framework and assumptions}

To begin is convenient to describe the framework in which we work and to give the main
definitions of the processes and variables involved. The framework setting follows strictly the
\emph{reduced-form models} literature and we refer to the classical monograph of \cite{biel04} for details.\\
Let us consider a probability space described by the triple $(\Omega, \mathcal{G}_{t}, \mathbb{P})$ where the full filtration is given by  $\mathcal{G}_{t} = \sigma(\mathcal{F}_{t} \vee \mathcal{H}^{A}\vee\mathcal{H}^{B})_{t\geq0}$ for $t\in[0,T]$  and $\mathbb{P}$  the real probability measure defined on this space. On it lie two strictly positive random time $\tau_{i}$ for $i\in \{A,B\}$, which represent the \emph{default times} of the counterparties considered in our model.  In addition, we define the \emph{default process } $H^{i}_{t} = \mathbbm{1}_{\{\tau^{i} \leq t\}}$ and the relative filtration $\mathcal{H}^{i}$ generated by $H_{t}^{i}$ for any $t\in \mathbb{R}^{+}$. We are left to mention $\mathcal{F}$ which is the (risk-free) \emph{market filtration} generated by a $d$-dimensional Brownian motion vector $W$  adapted to it, under the real measure $\mathbb{P}$. In addition, we remember that all the processes we consider, in particular $H^{i}$,  are \emph{c\'adl\'ag semimartingales} $\mathcal{G}$ adapted and $\tau^{i}$ are $\mathcal{G}$ stopping times.\\ 
For convenience, let us define the first default time of the counterparties as $\tau = \tau_{A}\wedge\tau_{B}  $ which also represent the ending/exstinction time of the underlying contract, with the corresponding indicator process $H_{t} = \mathbbm{1}_{\{\tau\leq t\}}$.
As concerns the underlying market model it is assumed arbitrage-free, namely it admits a \emph{spot martingale measure} $\mathbb{Q}$ (not necessarily unique) equivalent to $\mathbb{P}$. A spot martingale measure is associated
with the choice of the savings account $B_{t}$ (so that  $B^{-1}$ as discount factor) as a numeraire that, as usual,  is given by $\mathcal{F}_{t}$-predictable process
\begin{equation}
    B_{t}= \exp \int_{0}^{t}r_{s}ds, \:\:\forall \: t \in \mathbb{R}^{+}
\end{equation}
where the short-term $r$ is assumed to follow an $\mathcal{F}$-progressively measurable stochastic process (whatever it  is the choice of the term structure model for itself).\\
We then define the \emph{Az\'ema supermartingale} $G_{t} = \mathbb{P}(\tau > t |\mathcal{F}_{t})$ with $G_{0}=1$ and $G_{t} >0\:\: \forall \: t \in \mathbb{R}^{+} $  as the \emph{survival process} of the default time $\tau$ with respect to the filtration $ \mathbb{F}$. 
The process $G$ being a bounded $\mathcal{G}$-supermartingale it admits a unique \emph{Doob-Meyer decomposition} $G = \mu - \nu$, where $\mu$ is the martingale part and $  \nu$ is a predictable increasing process. In particular, $\nu$ is assumed to be absolutely continuous  with respect to the \emph{Lebesgue measure}, so  that $d\nu_{t} = \upsilon_{t} dt$ for some $\mathcal{F}$-progressively measurable, non-negative process $\upsilon$.
So that, we can define now the default intensity $\lambda$ as the $\mathcal{F}$-progressively measurable process that is set as $\lambda_{t} = G_{t}^{-1}\upsilon_{t}$ so that $dG_{t} = d\mu_{t} - \lambda_{t} G_{t}dt$ and the cumulative default intensity is defined as follows
\begin{equation}
    \Lambda_{t}=\int_{0}^{t} G_{u}^{-1} d\nu_{u} = \int_{0}^{t} \lambda_{u}du,
\end{equation}
For convenience, we assume that the \emph{immersion property} holds in our framework, so that every c\'adl\'ag $\mathcal{G}$-adapted (square-integrable) martingale is also $\mathcal{F}$-adapted. 


In particular, we assume \emph{pre-default valued processes} namely, setting $J := 1-H = \mathbbm{1}_{\{t\leq \tau \}}$ we have that for any $\mathcal{G}$-adapted, respectively $\mathcal{G}$-predictable process $X$ over $[0,T]$, there exists a unique $\mathcal{F}$-adapted,  respectively $\mathcal{F}$-predictable,  process $\tilde{X}$ over $[0,T]$, called the pre-default value process of $X$, such that $JX = J\tilde{X}$, respectively $J_{-}X = J_{-}\tilde{X}$.\\
As regards counterparty objectives, $\{A,B\}$ are both defaultable and are assumed to behave
rationally with the same objective to minimize the overall costs  related to counterparty risk -
quantified through the BCVA - and those related to collateral and funding. The information flow is assumed symmetric.

\subsection{\large Main definitions: BCVA, contingent CSA and funding}
Let us state the main definitions that are needed to model our claim with contingent CSA. Also
here we just state the objects involved, for proofs and details we refer to the already mentioned
work and the reference therein.\\

\textbf{1) BCVA definition}. Following \cite{biel11} (proposition 2.9), the bilateral CVA process of a defaultable claim with bilateral counterparty risk $(X; \mathbf{A};Z; \tau )$ maturing in $T$ satisfies the following relation\footnote{The formulation is seen from the point of view of $B$. Being symmetrical between the party, just the signs change.}
      \begin{eqnarray}
  BCVA_{t} &=& S_{t}^{rf} - S_{t} \\
  &=& CVA_{t} - DVA_{t} \nonumber \\
  &=& B_{t}\mathbb{E}_{\mathbb{Q}^{*}} \bigg[ \mathbbm{1}_{\{t< \tau=\tau_{B}\leq T\}}B_{\tau}^{-1}(1-R_{c}^{B})(S^{rf}_{\tau})^{-} \bigg|\mathcal{G}_{t} \bigg] + \nonumber \\
  &-& B_{t}\mathbb{E}_{\mathbb{Q}^{*}}\bigg[ \mathbbm{1}_{\{t< \tau=\tau_{A}\leq T\}}B_{\tau}^{-1}(1-R_{c}^{A})(S^{rf}_{\tau})^{+}\bigg|\mathcal{G}_{t}  \bigg]
\end{eqnarray}
for every $t \in[0,T]$, where $B_{t}$ indicates the \emph{discount factor},$R_{c}^{i}$ for $i\in\{A,B\}$ is the \emph{counterparty recovery rate }(process) and where the \emph{clean price process} $S^{rf}_{t}$  would be simply
represented by the integral over time of the contract dividend flow under the relative martingale pricing measure $\mathbb{Q}$, that is
\begin{equation}
   S_{t}^{rf}= B_{t}\mathbb{E}_{\mathbb{Q}} \bigg( \int_{]t,T]} B_{u}^{-1} dD^{rf}_{u} \big| \mathcal{F}_{t} \bigg)\;\;t\in[0,T]
\end{equation}
and $D^{rf}_{t}$ is the clean dividend process of the default-free contract
\begin{eqnarray}
  D_{t}^{rf} &=& X(T) + \sum_{i\in\{A,B\}}\bigg(\int_{]t, T]} d\mathbf{A}^{i}_{u} \bigg)  \;\;t\in[0,T]
\end{eqnarray}
where   $X$ is the $\mathcal{F}$-adapted final payoff, $\mathbf{A}$ the $\mathcal{F}$-adapted process representing the contract's \emph{promised dividends} and $\tau  =\tau^{i} = \infty$. We are left to state the definitions of (bilateral) \emph{risky dividend} and \emph{price process} of a general defaultable claim:
\begin{eqnarray}
  D_{t} &=& X \mathbbm{1}_{\{T<\tau\}}  + \sum_{i\in\{A,B\}}\bigg(\int_{]t, T]}  (1-H_{u}^{i})d\mathbf{A}^{i}_{u} +  \int_{]t,T] } Z_{u}dH_{u}^{i}\bigg)  \;\;t\in[0,T]
\end{eqnarray}
for $i \in\{A,B\} $ and
\begin{equation}
   NPV_{t} = S_{t}= B_{t}\mathbb{E}_{\mathbb{Q}} \bigg( \int_{]t,T]} B_{u}^{-1} dD_{u} \big| \mathcal{G}_{t} \bigg)\;\;t\in[0,T].\\
\end{equation}
where $Z$ is the  \emph{recovery process} that specifies the recovery payoff at default and $H_{t} := \mathbbm{1}_{\{ \tau \leq t \}}$ the already defined \emph{default process}.\\

\textbf{2)  Collateral definition with contingent CSA.} In order to generalize collateralization in
presence of contingence CSA, we recall the definition collateral account/process $Coll_{t}:[0,T]\rightarrow \mathbb{R}$ is a stochastic $\mathcal{F}_{t}$-adapted process defined as
\begin{equation}
    Coll_{t}=\mathbbm{1}_{\{S^{rf}_{t} > \Gamma_{B} + MTA\}}(S^{rf}_{t} - \Gamma_{B}) + \mathbbm{1}_{\{S^{rf}_{t} < \Gamma_{A} - MTA\}}(S^{rf}_{t} - \Gamma_{A}),
\end{equation}
on the time set $ \{t<\tau\}$, and
    \begin{equation}
    Coll_{t}=\mathbbm{1}_{\{S^{rf}_{\tau^{-}} > \Gamma_{B} + MTA\}}(S^{rf}_{\tau^{-}} - \Gamma_{B}) + \mathbbm{1}_{\{S^{rf}_{\tau^{-}} < \Gamma_{A} - MTA\}}(S^{rf}_{\tau^{-}} - \Gamma_{A}), \;
\end{equation}
 on the set $ \{\tau\leq t <\tau +\Delta t\}$, thresholds $\Gamma_{i}$, for $i \in \{A,B\}$ and positive minimum transfer amount $MTA$. \\
 The perfect collateralization case, say $Coll^{Perf}_{t}$, can be shown to be always equal to the
mark to market, namely to the (default free) price process $S^{rf}_{t}$ of the underlying claim,
that is formally
\begin{equation}
    Coll_{t}^{Perf}=\mathbbm{1}_{\{S^{rf}_{t} > 0 \}}(S^{rf}_{t} - 0) + \mathbbm{1}_{\{S_{t}^{rf} <0\}}(S^{rf}_{t} - 0) = S^{rf}_{t} \;\; \forall \:  t \in [0,T], \; on\:\{t<\tau\}.
\end{equation}
and
\begin{equation}
    Coll_{t}^{Perf}=  S^{rf}_{\tau^{-}} \;\; \forall \:  t \in [0,T], on \: \{ \tau\leq t <\tau+\Delta t\}.
\end{equation}
Let us remind that in presence of \emph{perfect/full collateralization} one can easily show that
\begin{eqnarray}
  BCVA_{t}^{Coll^{Perf}} &=& S_{t}^{rf} - S_{t} =0\Longrightarrow \nonumber \\
   S_{t}&=&  S_{t}^{rf}  \qquad \forall t\in [0,T]
\end{eqnarray}
Generalizing, the contingent collateral $Coll^{C}_{t}$ can be defined as the $\mathcal{F}_{t}$-adapted process
defined for any time $t\in [0, T]$ and for every switching time $\tau_{j} \in [0; T]$ and $j = 1,\dots,M$,
switching indicator $z_{j} $and default time $\tau$ (defined above as $\min\{\tau_{A}, \tau_{B}\} $), which is formally
\begin{equation}
    Coll^{C}_{t} = S^{rf}_{t}\mathbbm{1}_{\{ z_{j}=0 \}} \mathbbm{1}_{\{ \tau_{j} \leq t < \tau_{j+1} \}} + 0 \mathbbm{1}_{\{ z_{j}=1 \}} \mathbbm{1}_{\{ \tau_{j} \leq t < \tau_{j+1} \}} \;\: on \: \{ t<\tau\}
\end{equation}
on the set $ \{ t<\tau\}$, and
\begin{equation*}
    Coll^{C}_{t} = S^{rf}_{\tau^{-}}\mathbbm{1}_{\{ z_{j}=0 \}} \mathbbm{1}_{\{ \tau_{j} \leq t < \tau_{j+1} \}} + 0 \mathbbm{1}_{\{ z_{j}=1 \}} \mathbbm{1}_{\{ \tau_{j} \leq t < \tau_{j+1} \}} \;
\end{equation*}
on the set  $\{ \tau\leq t <\tau+\Delta t\}$.\\

 \textbf{2) BCVA definition with contingent CSA.} By the definition of $D_{t}^{C}$ and
$S_{t}^{C}$ as the dividend and price process in presence of the \emph{contingent CSA}
 \begin{eqnarray}
  D^{C}_{t} &=&  D^{rf}_{t}\mathbbm{1}_{\{ z_{j}=0 \}} \mathbbm{1}_{\{ \tau_{j} \leq t < \tau_{j+1} \}} + D_{t} \mathbbm{1}_{\{ z_{j}=1 \}} \mathbbm{1}_{\{ \tau_{j} \leq t < \tau_{j+1} \}} \\
  S^{C}_{t} &=&  S^{rf}_{t}\mathbbm{1}_{\{ z_{j}=0 \}} \mathbbm{1}_{\{ \tau_{j} \leq t < \tau_{j+1} \}} + S_{t} \mathbbm{1}_{\{ z_{j}=1 \}} \mathbbm{1}_{\{ \tau_{j} \leq t < \tau_{j+1} \}}
\end{eqnarray}
 for any time $t \in [0, T\wedge \tau]$, switching times $\tau_{j} \in  [0, T\wedge \tau]$,  $j = 1,\dots,M$ and  switching indicator $z_{j} \in \{0,1\}$, we define the bilateral CVA of a contract with contingent CSA of switching type as follows
 \begin{eqnarray}
  BCVA^{C}_{t} &=& S^{rf}_{t} - S^{C}_{t} \nonumber\\
   &=& S^{rf}_{t} - (S^{rf}_{t}\mathbbm{1}_{\{ z_{j}=0 \}} \mathbbm{1}_{\{ \tau_{j} \leq t < \tau_{j+1} \}} + S_{t} \mathbbm{1}_{\{ z_{j}=1 \}} \mathbbm{1}_{\{ \tau_{j} \leq t < \tau_{j+1} \}}) \nonumber\\
   &=&  0 \mathbbm{1}_{\{ z_{j}=0 \}} \mathbbm{1}_{\{ \tau_{j} \leq t < \tau_{j+1} \}} + BCVA_{t}\mathbbm{1}_{\{ z_{j}=1 \}} \mathbbm{1}_{\{ \tau_{j} \leq t < \tau_{j+1} \}}.
\end{eqnarray}
where the expression for $BCVA_{t}$ is known from the former point.\\

\textbf{4) Funding definition}. As regards funding, in our setting we allow for difference in the funding rates between counterparties. In particular, we assume the existence of the following
funding asset $B^{opp^{i}}_{t}$, $B^{borr^{i}}_{t}$ and $B^{rem^{i}}_{t}$
In particular, under the assumptions of \emph{segregation}
(no collateral \emph{rehypotecation}), collateral made up by cash and BCVA not funded\footnote{This assumption is relevant in order to simplify the problem formulation and to deal with its recursive nature.}, let us
highlight that if counterparty $i \in \{A,B\}$ has to post collateral in the margin account, she
sustains a funding cost, applied by the external funder, represented by the \emph{borrowing rate}
$r^{borr^{i}}_{t} = r_{t} + s_{t}^{i}$ that is the risk free rate plus a credit spread  (that is usually different
from the other party) . By the other side, the counterparty receives by the funder the remuneration on the collateral post, that we define in the CSA as a risk free rate plus
some basis points,  namely is $r_{t}^{rem^{i}} = r_{t} +bp_{t}^{i}$.
Hence we assume the following dynamics for the   funding assets (which can be different
between counterparties)
\begin{eqnarray}
         dB^{borr^{i}}_{t} &=& (r_{t} + s_{t}^{i})B^{borr^{i}}_{t}dt, \qquad i\; \in \{A,B\}  \\
        dB^{rem^{i}}_{t} &=& (r_{t}+bp_{t}^{i} )B^{rem^{i}}_{t}dt,\qquad i \;\in \{A,B\}
      \end{eqnarray}
Instead, considering the counterparty that call the collateral, as above the collateral is
remunerated at the rate given (as the remuneration for the two parties can be different)
by $B^{rem}$, but she cannot use or invest the collateral amount (that is segregated), so she
sustains an opportunity cost, that can be represented by the rate $r_{t}^{opp^{i}} = r_{t} + \pi_{t}^{i}$
where $\pi$ is a premium over the risk free rate. Hence, we assume the existence of the following asset
too
\begin{eqnarray}
         dB^{opp^{i}}_{t} &=& (r_{t} +\pi_{t}^{i})B^{borr^{i}}_{t}dt, \qquad i\; \in \{A,B\}
      \end{eqnarray}

To conclude the section, let us underline that given the symmetrical nature of processes (except for the funding ones) the following relations states:
\begin{eqnarray*}
  BCVA_{t}^{A} &=& -BCVA_{t}^{B}\qquad t \;\in [0,\tau\wedge T] \\
  BCVA_{t}^{C^{A}} &=& -BCVA_{t}^{C^{B}}\qquad t \;\in [0,\tau\wedge T]\\
  Coll_{t}^{C^{A}} &=& -Coll_{t}^{C^{B}} \qquad t \;\in [0,\tau\wedge T].
\end{eqnarray*}

\subsection{\large Model dynamics, controls and cost functionals}
In our contingent CSA model of multiple switching type (with finite horizon), both counterparties
$A,B$ are free to switch from zero to perfect collateralization every time in $[0, T]$.  Hence their
control sets are made up by sequences of \emph{switching times} - say $\tau_{j}\in \mathcal{T}$  - and \emph{switching indicators} $z_{j} \in  \mathcal{Z}$ with $\mathcal{T} \subset [0,T]$, that we define formally as follows
\begin{equation}
    \mathcal{C}^{A}=\big\{ \mathcal{T}^{A}, \mathcal{Z}^{A} \big\} = \big\{ \tau_{j}^{A}, z_{j}^{A}\big\}_{j=1}^{M}, \; \forall \tau_{j}^{A} \in [0,T],\; z_{j}^{A} \in \{0,1\}
\end{equation}
  \begin{equation}
    \mathcal{C}^{B}=\big\{ \mathcal{T}^{B}, \mathcal{Z}^{B} \big\} = \big\{ \tau_{j}^{B}, z_{j}^{B}\big\}_{j=1}^{M}, \; \forall \tau_{j}^{B} \in [0,T],\; z_{j}^{B} \in \{0,1\}
\end{equation}
with the last switching $\{\tau^{M}_{i}\leq T  \}$ $(M<\infty)$. We recall that $\tau_{j}^{i}$ are by definitions of \emph{stopping times} $\mathcal{F}_{t}$-measurable random variables, while $z_{j}^{i}$ are $\mathcal{F}_{\tau_{j}}$-measurable switching indicators, taking values in our model $\forall \j \in 1\,\dots,M$
\[
\begin{cases}
z_{j} = 1 \Rightarrow   & \emph{zero collateral}\; (full \;CVA)\\
z_{j} = 0 \Rightarrow   & \emph{full collateral}\; (null \;CVA)\\
\end{cases}
\]
Clearly controls affect also our model dynamic. As regards this point, we assume general
markovian diffusions for $(X; \lambda^{i})$ $i\in \{ A,B\}$,   namely the interest rates and the default intensities of counterparties. From the definitions of \emph{contingent CSA} and (bilateral) CVA given in
the last section, we highlight that the switching controls enter and affect the dynamic of these
processes. In fact, as we know, switching to full collateralization implies $BCVA = 0$, that is
$S_{t} = S^{rf}_{t} =Coll_{t}^{perf}$. This means no counterparty risk and so the default intensities' dynamic
$d\lambda_{t}^{i}$  won't be relevant, just $dX$ will be considered while $z = 0$ ( namely until the
collateralization will be kept active). So, formally, we have:\\

$if \: \big\{z_{j}= 1\big \} \: and \: \{\tau_{j}\leq t<\tau_{j+1}\}\Rightarrow \;$
\begin{eqnarray*}
  D_{t}^{C} &=& D_{t} \Rightarrow\\
  S_{t}^{C} &=& S_{t} \Rightarrow\\
  BCVA_{t}^{C} &=& BCVA_{t} \forall\:t\in[0,T\wedge \tau]
\end{eqnarray*}
so that the relevant dynamic to model the BCVA process in this regime is
\begin{eqnarray*}
  dX_{t} &=& \mu(t,X_{t})X(t) dt + \sigma(t,X_{t}) X(t) dW_{t}^{x};\quad\qquad X(0)=x_{0} \\
  d\lambda_{t}^{A} &=& \gamma(t,\lambda_{t}^{A})\lambda^{A}(t)dt + \nu(t,\lambda_{t}^{A}) \lambda^{A}(t)dW_{t}^{\lambda^{A}}; \qquad \lambda^{A}(0)=\lambda^{A}_{0}\\
  d\lambda_{t}^{B} &=& \chi(t,\lambda_{t}^{B})\lambda^{B}(t)dt + \eta(t,\lambda_{t}^{B}) \lambda^{B}(t)dW_{t}^{\lambda^{B}}; \qquad \lambda^{B}(0)=\lambda^{B}_{0}\\
 d\langle X,\lambda_{t}^{A}\rangle_{t} &=& d\langle W_{t}^{x},W_{t}^{\lambda^{A}}\rangle_{t} = \rho_{X,\lambda^{A}}dt\\
 d\langle X,\lambda_{t}^{B}\rangle_{t} &=& d\langle W_{t}^{x},W_{t}^{\lambda^{B}}\rangle_{t} = \rho_{X,\lambda^{B}}dt\\
\end{eqnarray*}

$if \: \big\{z_{j}= 0 \big \} \; and \; \{\tau_{j}\leq t<\tau_{j+1}\}\Rightarrow $
\begin{eqnarray*}
  D_{t}^{C} &=& D_{t}^{rf} \Rightarrow\\
  S_{t}^{C} &=& S_{t}^{rf} =Coll_{t}^{Perf}  \Rightarrow\\
  BCVA_{t}^{C} &=& 0 \forall\:t\in[0,T\wedge \tau]
\end{eqnarray*}

so that the relevant dynamic  to model the process in this regime will be just
\begin{equation*}
  dX_{t} = \mu(t,X_{t})X(t) dt + \sigma(t,X_{t}) X(t) dW_{t}^{x};\quad X(0)=x_{0}.
\end{equation*}
Here, the drift and volatility coefficient $\mu(t,x)$, $\sigma(t,x)$, $\gamma(t,x)$, $\nu(t,x)$, $\chi(t,x)$ and $\eta(t,x)$ are all continuous, measurable and real valued function $\mathcal{F}$-adapted to the relative brownian filtration.
For a matter of convenience we ease the notation by setting our system dynamic in vectorial form;\\
\begin{center}
$d\mathcal{Y}(t):=\left[
  \begin{array}{c}
     dt\\
    dX_{t} \\
    d\lambda^{i}_{t} \\
    dZ^{i} \\
  \end{array}
\right]$, $ \qquad \mathcal{Y}(0)=\left[
  \begin{array}{c}
     t=0\\
     x_{0} \\
    \lambda^{i}_{0} \\
    Z_{0}^{i}=1 \\
  \end{array}
\right ]$
\end{center}
for $i \in \{A,B\}$. \\

As regards counterparties cost functionals' formulation, we remind that both are assumed
coherently\footnote{Otherwise, it would not have sense for both to sign a contract wich give
flexibility to activate collateralization whenever is optimal.} counterparty risk averse, but in this case they can have different preference/cost functions in which they need to take in account also the optimal control strategy of the other party, namely its response function $b^{-i}(u^{i})$  where $u_{i} := \mathcal{C}^{i}$ and $i\in  \{A,B\}$. We are going to discuss more about it in the next section on game formulation. Here, let us be more explicit
about the formulation of counterparties costs for which we assume - for convenience - quadratic
preferences for both  generalized to take in account the optimal control strategy of the
other party over time $t \in [0; \tau \wedge T]$, that is formally\footnote{For further details let us refer to section three of Mottola (2013).}:\\

\textbf{a)}  \textbf{Running costs}:
   \[
 F^{A}(\mathcal{Y}_{t},b^{B}(u^{A}),t) = \begin{cases}
        \big[(CVA^{A}(s)-DVA^{A}(s)) - \delta^{B}(s,u^{*,A}) \big]^{2} & if \{z_{j}^{i}=1\}\\
       \Big( \big( \int_{u}^{T\wedge\tau^{A}_{j+1}}  R^{A}(s) [NPV^{A}(v)]du - NPV^{A}(s)\big)- \delta^{B}(s,u^{*,A}) \Big)^{2} & if\; \{z_{j}^{i}=0\} .
 \end{cases}\]
 $\forall \; s, \tau_{j}\in[t,T\wedge\tau]$ and for counterparty $B$
   \[
 F^{B}(\mathcal{Y}_{t},b^{A}(u^{B}),t) = \begin{cases}
        \big[(CVA^{B}(s)-DVA^{B}(s)) - \delta^{A}(s,u^{*,B}) \big]^{2}  & if \{z_{j}^{i}=1\}\\
       \Big( \big( \int_{u}^{T\wedge\tau^{B}_{j+1}}  R^{B}(s) [NPV^{B}(v)]du - NPV^{B}(s)\big)- \delta^{A}(s,u^{*,B}) \Big)^{2} & if\; \{z_{j}^{i}=0\} .
 \end{cases}
 \]
 $\forall \; s, \tau_{j}\in[t,T\wedge\tau]$ and $i\in  \{A,B\}$. Here, all the terms of the running costs are known except the response function $\delta(.)^{i}$ which is assumed non-negative, continuous and $\mathcal{F}$-adapted and
 the funding factor term $R^{i}(s)$ which we set to model the expected collateral and funding costs when $z_{j}=0$, that is formally

\[
 R^{i}(t) = \begin{cases}
        -\exp-(r_{borr}^{i} - r_{rem}^{i})t  & if \; z_{j}=0 \;and \;NPV<0 \\
        \exp-(r_{opp}^{i} - r_{rem}^{i})t & if \;z_{j}=0 \;and \;NPV>0.
        \end{cases}
\]

 \textbf{b)} \textbf{Terminal costs}
   \[
 G^{A}(\mathcal{Y}_{t},b^{B}(u^{A}),t) =
 \begin{cases}
       (-NPV^{A}(T)-\delta^{B}(T,u^{*,A}))^{2} \Rightarrow& if \; collateral \; is\; active  \\
       ( 0-\delta^{B}(T,u^{*,A}))^{2}\Rightarrow  & \;no  \;collateral
 \end{cases}\]

   \[
 G^{B}(\mathcal{Y}_{t},b^{A}(u^{B}),t) =
 \begin{cases}
       (-NPV^{B}(T)-\delta^{A}(T,u^{*,B}))^{2} \Rightarrow & if \; collateral \; is\; active \\
       ( 0-\delta^{A}(T,u^{*,B}))^{2}\Rightarrow  & no  \;collateral.
 \end{cases}
 \]

\textbf{c)} \textbf{Instantaneous switching costs}:
   \begin{equation*}
        l^{i}\big( \tau_{j}^{i} , z_{j}^{i} \big) = \sum_{j\geq  1}^{M}e^{-r\tau_{j}^{i}} c_{z_{j}^{i}}(t)\mathbbm{1}_{\{ \tau_{j}^{i} \wedge \tau_{j}^{-i}<T\}} , \;\: \forall\:\:\tau_{j}^{i},z_{j}^{i}  \in \{\mathcal{T}^{i}, \mathcal{Z}^{i} \},
    \end{equation*}
    for $i \in \{A,B\}$, where $c^{i}(.)$ is the $\mathcal{F}$-predictable (deterministic for convenience)  instantaneous cost function.

\subsection{\large Game formulation and pure strategies definition}

In this section we pass to give a generalized formulation for our contingent CSA scheme in which
allowing for the strategic interaction between the players - which are the counterparties of this
theoretical contract - we are lead to formulate our problem as a stochastic differential game
whose study of the equilibrium is central for the existence of a solution and the optimal design
of our contingent scheme.\\
In order to formulate the game, firstly, we recall that in our model for the contingent CSA
scheme we assume no fixed times or other rules for switching, that is the counterparty can switch
optimally every time until contract maturity T in order to minimize its objective functional. But
the functionals, as set formally in the former section, now are generalized and not symmetrical
between the parties : as already mentioned, both players are assumed to remain \emph{risk averse}
to the variance of bilateral CVA, collateral and funding costs, but depending on the different
parametrization of the functionals (that we show below) and instantaneous switching costs other
than the difference in default intensities, the problem can be naturally represented by a generalized \emph{non-zero-sum Dynkin game}. This is a non-zero-sum game given that the player payoff
functionals are not symmetrical and generalized in the sense that player controls are not just
simple \emph{stopping times} but sequences of random times that define the optimal times to switch $\tau_{j}$ from
a regime to the other (together with switching indicators sequences $z_{j}$ ).\\
Therefore, given that the \emph{right to switch} is bilateral and we assume no other rules/constraint
on controls set by contract, the other player's optimal strategy - and hence the strategic interaction with the other party - becomes central to define the own optimal switching strategy.
Let us be more formal and, building on the definitions of section 2.3, we define our model's SDG
as a generalized Dynkin game of switching type as follows.\\

\textbf{Definition 2.4.1 (Dynkin game of switching type definition)}.\emph{Let us consider two
players/counterparties $\{A,B\}$ that have signed a general contract with a contingent CSA of
switching type. Given the respective payoff functionals $F^{i}(.)$
(or running reward), the terminal rewards $G^{i}(.)$ and the instantaneous switching cost functions $l^{i}(.)$ where $i \in \{A,B\}$, under rationality assumption and non-cooperative strategic interaction, the players aim to minimize the following objective functional:
\begin{eqnarray}
    J^{i}(y,u^{i},u^{-i})&=& \inf_{ u^{i}  \in\{\mathcal{C}^{i}_{ad}\}} \mathbb{E} \Bigg[\sum_{j} \int_{t}^{\tau_{j}^{i}\wedge\tau_{j}^{-i}\wedge T} B_{s}\bigg[F^{i}(y_{s},u^{i}, b^{-i}(u^{i}))\bigg]ds \\ \nonumber
      &+&  l^{i}\big( \tau_{j}^{i}, z_{j}^{i} \big)   + G^{i}(y_{T}, b^{-i}(u^{i})) \bigg| \mathcal{F}_{t} \Bigg]\;\; for \:i \in\{A,B\}
\end{eqnarray}
where we mean for $i = A$ then $-i = B$ and viceversa, $B_{t}$ is defined in (1),  the system dynamic is defined  in section 2.3 by $dy_{t}:= d\mathcal{Y}_{t}$, the controls set are defined in (21)-(22) and we have set for notational convenience $u_{i} :=\{\mathcal{T}^{i},\mathcal{Z}^{i}\}$ for $ i \in\{ A,B\}$.}\\

Let us  underline from definition 2.4.1 that  the payoffs functions, whose specific formulation has been stated in section 2.3, can differ  between $A$ and $B$ for the following terms
\begin{eqnarray}
  \delta^{A}(.) &\lesseqgtr& \delta^{B}(.) \\
  R^{A}_{t}(.) &\lesseqgtr& R^{B}_{t}(.) \\
  c^{A}_{t}(.) &\lesseqgtr& c^{B}_{t}(.)
\end{eqnarray}
where\\
\noindent a) $R^{i}(.)$ is the funding-collateral cost factor defined in the former section;\\
b) $\delta^{i}(.)$ is the running cost function threshold which is generalized her taking as argument the
optimal response and control strategies of the other player;\\
c) $c^{i}_{t}(.)$ are the already known instantaneous costs from switching.\\

\textbf{Remarks 2.4.2.} The game as formulated above in (23) is fairly general; in addition,   one could also introduce the possibility for the players to stop the game adding a stopping time (and the related reward/cost function) to the set of controls made up of switching times and indicators. From the financial point of view, this can be justified by a \emph{early termination clause} set in the contingent CSA defined by the parties. Anyway, given the problem recursion, this would add  greater complications that we leave for further research.\\
Actually this game is already complicated by the fact that, differently from the  (non-zero-sum) Dynkin game as formulated in section 1.3, here the players control strategies affect  also each other   payoffs. In fact, given that in our general formulation the players can switch optimally whenever over the life of the underlying contract, it is clear that - without setting any other \emph{rules} for the game - the decision of one player to switch to a certain regime impose a different cost function $F_{Z}(.)^{i}$ also for the other player. So if $A$ switches but for $B$ the decision is not optimal, he is able to immediately switch back, taking in account the instantaneous switching costs\footnote{Note from (23) that  the indicator $\mathbbm{1}_{\{\tau_{j}^{A}\wedge \tau_{j}^{B}<T \}}$ the instantaneous switching costs enter in the functional whoever of the players decides to switch}.
In this sense, the relative difference between players' payoff  (in the different regimes) and the strategic interaction between them over time become central in order to understand and analyze the problem solution/equilibrium. \\

We return on this points later, here is important to mention that in order to highlight this strategic dependence in the game - that is assumed to be played  by \emph{rational  and non-cooperative} players - we have enriched  the running cost function $F_{Z}(.)^{i}$ by a response function $b^{-i}(u^{i})$, which can be intended mainly in two way:\\

\indent a) \emph{as the ''classical''  best response function to the other player strategy, which implies the complete information assumption in the game, that is the players have the same information set about the system dynamic and they are able to calculate (under the real probability measure $\mathbb{P}$) each other payoff;}\\
\indent b) \emph{if the game information is not complete and there is a degree of \emph{uncertainty} over the players payoff and their switching strategy, the function $b^{-i}(u^{i})$ can be intended in generalized terms as a probability distribution assigned by  a player to the optimal response of the other one.}\\

We discuss further on the game information flows below. Now, the main issue to tackle is to understand the  condition under which this generalized game (23)  have sense and it will be played, which means that it will be signed by counterparties. This takes to the  problem definition of an equilibrium for this game  and to the condition under which its existence and uniqueness are ensured.\\
Before giving the formal definition of the game equilibrium, let us highlight the game pure strategies at a given time $ \{\tau_{j-1}^{i}< t\leq \tau_{j}^{i}\}$ under the assumption of \emph{simultaneous moves} by players.\\\\

\textbf{Definition 2.4.3 (Pure strategies of the game of switching type).} \emph{For any given initial condition ${z_{0}^{A},z_{0}^{B} }$ and $\forall\: z_{j}^{A} \in u^{A}$ and $z_{j}^{B} \in u^{B}$ and $ \{\tau_{j-1}^{i}< t\leq \tau_{j}^{i}\}$, the pure strategies of our Dynkin game of switching type are defined as follows\\
$if\{z_{j-1} = 0\}\:\Longrightarrow$
\begin{eqnarray*}
         \{z_{j}^{A}&=& 0, \; z_{j}^{B}= 0\}\: \Longrightarrow \: "no\: switch" \\
        \{z_{j}^{A} &=& 0, \; z_{j}^{B}=1\}  \: \Longrightarrow \: " switch\: to\: 1"  \\
        \{z_{j}^{A} &=& 1, \; z_{j}^{B}=0\} \: \Longrightarrow \: " switch\: to\: 1"\\
        \{z_{j}^{A} &=& 1, \; z_{j}^{B}=1\} \: \Longrightarrow \: " switch\: to\: 1".
\end{eqnarray*}
while if $\{z_{j-1} = 1\}\:\Longrightarrow$
 \begin{eqnarray*}
         \{z_{j}^{A}&=& 0, \; z_{j}^{B}= 0\}\: \Longrightarrow \: "switch\:to \:0" \\
        \{z_{j}^{A} &=& 0, \; z_{j}^{B}=1\}  \: \Longrightarrow \: " switch\: to\: 0"  \\
        \{z_{j}^{A} &=& 1, \; z_{j}^{B}=0\} \: \Longrightarrow \: " switch\: to\: 0"\\
        \{z_{j}^{A} &=& 1, \; z_{j}^{B}=1\} \: \Longrightarrow \: " no \:switch".
\end{eqnarray*}}

\noindent In the table below we represent the standard game form at a given decision time with the possible (pure) strategies (namely the switching indicators) and the related random payoff between parenthesis.\\
      \begin{center}
      \begin{tabular}{|c|c|c|}
  \multicolumn{3}{}{}\\
  \hline
  $A,B$& $Switch$ & $No\: Switch$\\
  \hline
  $Switch$ & $1,1\:(J^{A},J^{B})$  & $1,0 \:(J^{A},J^{B})$  \\
  \hline
  $No\: Switch$ &$0, 1\:(J^{A},J^{B})$&  $0,0 \:(J^{A},J^{B})$\\
\hline
\end{tabular}
\end{center}

where we note that the players' strategies can be cast in these two categories:
\begin{enumerate}
  \item on the main diagonal of the table we have \emph{accomodation/peace type switching strategies} played
over time;
  \item on the opposite diagonal of the table we have \emph{fighting/war type switching strategies } played
over time.
\end{enumerate}

\subsection{\large Game equilibrium and stochastic representation through system of RBSDE }

From a static point of view the NEP for the game of definition 2.4.1 can be easily found once the payoff $J^{i}$ are known. But the problem is that  game configurations like these has to be played over time taking in account as key factors:
\begin{enumerate}
\item \emph{the payoff value that derives from switching at a given time;}
\item \emph{ the expected value from waiting until the next switching time;}
\item \emph{ the optimal responses, namely the other party optimal switching strategy. This implies - by information flows' symmetry - that each player  knows how to calculate (under the real probability measure $\mathbb{P}$) the  points a) and b) relative to the other party.}
\end{enumerate}

The resulting equilibrium is an optimal sequence of switching over time for both players that needs a (backward) dynamically recursive valuation. On an heuristic base, we expect that if the relative difference (over time) in players payoff functionals - mainly due to different function parametrization, default intensities $\lambda^{i}$ or switching costs $c_{Z}^{i}$ - remains low, it is more likely that the switching strategies on the main diagonal of the game $\{1,1;0,0\}$ will be played (given that both players would have also similar best responses). This should ease the search for the equilibrium of the game but this makes more likely to incur in \emph{banal solutions} . Otherwise, one should observe a more complicated strategic behavior that needs a careful study depending also on the type of equilibrium that now we try to define.\\
Indeed, given the characteristics of our game namely a non-zero-sum game in which the agents are assumed rational and act in a non-cooperative way in order to minimize their objective function knowing that also the other part will do the same,  the equilibrium/solution of this type of games is the celebrated \emph{Nash equilibrium point} (NEP). Actually, - as already shown - in our case the equilibrium  is  characterized by a sequence  of optimal switching over time and being game (23) a generalization of a \emph{Dynkin game}, by similarity, we can state the following definition of a \emph{Nash equilibrium point} for a \emph{Dynkin game of switching type}.\\\\

\textbf{Definition 2.5.1 (NEP for Dynkin game of switching type).} \emph{Let us define the switching control sets for the player $\{A,B\}$ of the generalized Dynkin game  (23) as follows
\begin{equation*}
    u_{A} :=   \big\{ \tau_{j}^{A}, z_{j}^{A}\big\}_{j=1}^{M}, \;\;\forall\: \tau^{A}_{j}\in[0,T],\:z^{A}_{j}\in\{0,1\};
\end{equation*}
\begin{equation*}
    u_{B} :=  \big\{ \tau_{j}^{B}, z_{j}^{B}\big\}_{j=1}^{M}, \;\;\forall\: \tau^{B}_{j}\in[0,T],\:z^{B}_{j}\in\{0,1\}.
\end{equation*}
A Nash equilibrium point for this game  is given by the pair of sequences of switching times and indicators $\{u^{*}_{A}, u^{*}_{B}\} $ such that for any control sequences $\{ u_{A}, u_{B}\}$ the following condition are satisfied
\begin{equation}
  J^{A}(y;u^{*}_{A}  ,u^{*}_{B} )  \leq J^{A}(y; u_{A}  ,u^{*}_{B} )
\end{equation}
and
\begin{equation}
J^{B}(y; u^{*}_{A}  ,u^{*}_{B} )  \leq J^{B}(y; u_{A}^{*}  ,u_{B} )
\end{equation}
(the signs will be reversed in the maximization case). }\\\\

\noindent A formal and rigorous  proof of the existence (and uniqueness) of a NEP for our  game such that it is non trivial or \emph{banal} in the sense that it is never optimal for both the parties to switch
or when the switching control set reduce to a single switching/stopping time - is the big issue
here.\\
In order to approach its solution , as we know from the introduction,  one has in general two way: analytic or probabilistic which are deeply interconnected (working in a markovian framework).
In particular, from the theory of BSDE with reflection\footnote{We refer in particular to the classical work of \cite{Elk97}}, we can state the next definition for the stochastic representation of our Dynkin game of switching type as a \emph{system of interconnected (non-linear) reflected BSDE}. Let us denote first with
\begin{equation*}
  \mathcal{M}^{p}=  \{ E[\sup_{t\leq T}|\nu_{t}|^{p}]<\infty \}
\end{equation*}
the set of progressively measurable and $p$ integrable processes $\nu_{t}$ and with
\begin{equation*}
\mathcal{K}^{p} = \{E[\int_{0}^{T}|\nu_{s}|^{p}ds] <\infty \}
\end{equation*}
the set of continuous and progressively measurable processes. Hence the following states.\\

\textbf{Definition 2.5.2 (RBSDE representation for game of switching type).}
\emph{Let us define the vector  triple $(Y^{i,Z}, N^{i,Z},K^{i,Z})$ for $i\in\{A,B\}$ and $Z\in\{z,\zeta\}$ with $Y^{i}$ and $N^{i}$ assumed progressively measurable and adapted to the market filtration $(\mathcal{F}_{t}^{W})$  and $K$ continuous and increasing. Then, given the standard Brownian motion vector $W_{t}$, the terminal reward $\xi^{i}$, the obstacles $Y^{i,Z}_{t} - c^{i,Z}_{t}$ and  the generator functions $F^{i}_{Z}(.,Y^{-i})$ assumed progressively measurable, uniformly Lipschitz and interconnected between the players, the Dynkin game of switching type formulated in (23) has  the following  representation through system of interconnected non-linear reflected BSDE}
\[
\begin{cases}
    Y^{A,z},Y^{A,\zeta}  \in \mathcal{K}^{2}; N^{A,z},N^{A,\zeta} \in \mathcal{M}^{2};\:\: K^{A,z}, K^{A,\zeta} \in \mathcal{K}^{2}, \: K \:non\:decreasing\:and\:K_{0} =0,\\
  Y_{t}^{A,Z} =\xi^{A}+ \int_{s}^{T} F^{A}_{Z}(y_{s},u^{A},N^{A}_{s};Y^{B}_{s})ds - \int_{s}^{T} N_{s}^{A,Z}dW_{s} + K^{A,Z}_{T} - K^{A,Z}_{s}, \; \: t\leq s\leq T\wedge \tau,\; Z\in\{z,\zeta\} \\
  Y_{t}^{A,z} \geq (Y_{t}^{A,\zeta} - c^{A,z}_{t}) ;\;\:  \int_{0}^{T}[ Y_{t}^{A,z} - (Y_{t}^{A,\zeta} - c^{A,z}_{t} )]dK^{A,z}_{t}=0;\\
  Y_{t}^{A,\zeta} \geq (Y_{t}^{A,z} - c^{A,\zeta}_{t}) ;\;\; \int_{0}^{T}[ Y_{t}^{A,\zeta} - (Y_{t}^{A,z} - c^{A,\zeta}_{t} )]dK^{A,\zeta}_{t}=0;\\\\
    Y^{B,z},Y^{B,\zeta}  \in \mathcal{K}^{2}; N^{B,z},N^{B,\zeta} \in \mathcal{M}^{2};\:\: K^{B,z}, K^{B,\zeta} \in \mathcal{K}^{2}, \: K \:non\:decreasing\:and\:K_{0} =0,\\
  Y_{t}^{B,Z} = \xi^{B}+ \int_{s}^{T} F^{B}_{Z}(y_{s},u^{B},N^{B}_{s};Y^{A}_{s})ds - \int_{s}^{T} N_{s}^{B,Z}dW_{s} + K^{B,Z}_{T} - K^{B,Z}_{s}, \; \: t\leq s\leq T\wedge \tau \;, Z\in\{z,\zeta\} \\
  Y_{t}^{B,z} \geq (Y_{t}^{B,\zeta} - c^{B,z}_{t}) ;\;\:  \int_{0}^{T}[ Y_{t}^{B,z} - (Y_{t}^{B,\zeta} - c^{B,z}_{t} )]dK^{B,z}_{t}=0;\\
  Y_{t}^{B,\zeta} \geq (Y_{t}^{B,z} - c^{B,\zeta}_{t}) ;\;\; \int_{0}^{T}[ Y_{t}^{B,\zeta} - (Y_{t}^{B,z} - c^{B,\zeta}_{t} )]dK^{B,\zeta}_{t}=0\\\\
\end{cases}
\]

From definition 2.5.2, it is evident that the system of RBSDE is a non-standard one given the characteristics of the generator functions  (which are the cost function in our game) that are inter-dependent and this is highlighted by the presence of the other player value process $Y^{i}_{t}$ inside $F^{i}_{Z}(.)$ for $i\in\{A,B\}$ . This makes hard to show the existence and uniqueness of the solution of the system for the reason that we highlight below. In particular, the solution of this system of RBSDE is made up of a two-dimensional vector made up by the triple $(Y^{*,Z}, N^{*,Z},K^{*,Z})$ where its dimension is given by the two switching regimes while the optimal switching sequences  is determined by the value process crossings of the barriers, represented  by the last two lines of the RBSDEs system.\\ Therefore, the other central issue is to show that the the system vector solution $Y^{*,Z}$   coincides with the players value functions'  of the non-zero-sum game of switching type (23).\\
As far as we know, these issues have been tackled - in relation to switching problems - in the already mentioned work of \cite{Ham10}. They study general system of m-dimensional BSDE called with \emph{oblique reflection}, which are RBSDE with both generator and barrier interconnected as in our case, showing existence and uniqueness of the solution while  the optimal strategy in general does not exist but an \emph{approximating optimal strategy} is constructed (through some technical estimates).\\

Let us briefly recall the main technical assumptions that are imposed in order to derive these results are:\\
\begin{itemize}
  \item \emph{square  integrability for both the generator function $F^{i}(.)$ and terminal reward $\xi$ while the obstacles function are continuous and bounded;}
  \item \emph{Lipschitz (uniform) continuity of  the generator function respect to its terms;}
  \item \emph{both the generator and the obstacles are assumed to be increasing function of the other players utility/value process.}
\end{itemize}

As also mentioned in the above mentioned paper, the condition c) implies from a game point of view that the players are \emph{partners}, namely the interaction and the impact of the other players value processes has a unique positive sign. This is not the case of our non-zero-sum game in which the interaction allowed between the two  players is antagonistic and more complicate.\\ 
Hence,  as far as we know, the existence and uniqueness solution of the optimal switching strategy for our
game formulated in definition 2.5.2 is an open problem, whose solution needs further studies.
Probably a solution for the problem exists but it won't be unique, indeed the classifications of solutions behavior'
and the conditions for their existence and uniqueness it is an interesting and hard program to tackle analytically
and also numerically.
Therefore, even though one could assume to simplify the problem in order to work under the same assumptions a)- c) that would ensure the existence and uniqueness of the solution, it would remain to verify that the solution of the system of RBSDE is the  \emph{Nash equilibrium point} for the game (23), which is complicated by the fact that the optimal control strategy may not exist\footnote{See \cite{Ham10} for details.}. Formally, one should prove the following theorem which is also an open problem.\\

\textbf{Theorem 2.5.3 (NEP and  RBSDE system solution).} \emph{Let us assume the existence and uniqueness of the solution for the system of definition 2.5.2, under the assumption a)- c). Then the system RBSDE value processes $Y^{*}_{A}$, $Y^{*}_{B}$ coincide with the player value functions of the game of switching type, that is
\begin{eqnarray*}
   Y^{*}_{A}&=& J^{A}(y,u^{*}_{A},u^{*}_{B})  \\
  Y^{*}_{B}&=& J^{B}(y,u^{*}_{B},u^{*}_{A})
\end{eqnarray*}
and are such that condition (27) and (28) are satisfied, which implies the existence and uniqueness of a Nash equilibrium point for the game (23).}\\

\textbf{Remarks 2.5.4.} As we already know, in the markovian framework - thanks to \cite{Elk97}
 results - the solution of the system of RBSDE is connected with the viscosity solution of a
generalized system of non-linear PDE with generator and obstacles different and interconnected
between the two players, which is even harder to study analytically. The main alternative is
to try to approach numerically the problem, searching for the conditions under which one can
find the equilibrium. A possibility is to apply the same technique - \emph{Snell envelope} and \emph{iterative
optimal stopping} technique of  the work of \cite{Car10} \footnote{We refer also to the fifth section of \cite{Mot13}}. adapted to study our stochastic game's solution. In particular, the algorithm need to be generalized in  order to introduce the players' strategic interaction and to compute the \emph{Nash equilibrium point } of the game.\\
Let us give here just a sketch of the numerical solution founded on the \emph{iterative optimal stopping approach} which is well suited to study the solution of our highly nonlinear and recursive problem. Let us  pick for exposition convenience two calculation times $t_{1}$ and $t_{2}$ and a final  regime switching  condition (given that the program run backward in time while the information grow forward) thanks to the \emph{dynamic programming principle}, both the players need to evaluate at these discretized times
\begin{enumerate}
  \item the optimality of an immediate switch at $t_{1}$ to the other regime ($Z \in \{z,\zeta\}$) taking in account the \emph{best response function} of the other player (over each switching time);
  \item the optimality to \emph{continue} namely to wait the next switching time $t_{2}$, considering also in this case the \emph{best response} of the other party.
\end{enumerate}
Formally, this means to  run the following program:
\begin{eqnarray}
          V^{l,A}(t_{1},\mathcal{Y}_{t_{1}},u^{A},b^{B}(u^{A}) ) &=& \min\bigg( \mathbb{E}\big[\int_{t_{1}}^{t_{2}} F^{A}(s,\mathcal{Y}_{s},u^{A}_{s},b^{B}(u^{A}_{s}) )ds + V^{l,A}(t_{2}, \mathcal{Y}_{t_{2}},u^{A}_{t_{2}},b^{B}(u^{A}_{t_{2}})) |\mathcal{F}_{t_{1}}\big], \nonumber\\
          & & SW^{A,Z}(t_{1},\mathcal{Y}_{t_{1}},u^{A}_{t_{1}},b^{B}(u^{A}_{t_{1}}) )\bigg)  \nonumber\\
           &\simeq& \min\bigg( F^{A,Z}(t_{1},\mathcal{Y}_{t_{1}},u^{A}_{t_{1}},b^{B}(u^{A}_{t_{1}}))\Delta t +  \mathbb{E}\big[  V^{l,A}(t_{2}, \mathcal{Y}_{t_{2}},u^{A}_{t_{2}},b^{B}(u^{A}_{t_{2}}))|\mathcal{F}_{t_{1}}\big], \nonumber\\
           & & \{V^{l-1, A}(t_{1},\mathcal{Y}_{t_{1}},u^{A}_{t_{1}},b^{B}(u^{A}_{t_{1}}))- c_{t_{1}}^{Z}\}\bigg)
\end{eqnarray}
\begin{eqnarray}
          V^{l,B}(t_{1},\mathcal{Y}_{t_{1}},u^{B},b^{A}(u^{B}) ) &=& \min\bigg( \mathbb{E}\big[\int_{t_{1}}^{t_{2}} F^{B}(s,\mathcal{Y}_{s},u^{B}_{s},b^{A}(u^{B}_{s}) )ds + V^{l,B}(t_{2}, \mathcal{Y}_{t_{2}},u^{B}_{t_{2}},b^{A}(u^{B}_{t_{2}})) |\mathcal{F}_{t_{1}}\big],\nonumber\\
          & & SW^{B,Z}(t_{1},\mathcal{Y}_{t_{1}},u^{B}_{t_{1}},b^{A}(u^{B}_{t_{1}}) )\bigg)  \nonumber\\
           &\simeq& \min\bigg( F^{B,Z}(t_{1},\mathcal{Y}_{t_{1}},u^{B}_{t_{1}},b^{A}(u^{B}_{t_{1}}))\Delta t +  \mathbb{E}\big[  V^{l,B}(t_{2}, \mathcal{Y}_{t_{2}},u^{B}_{t_{2}},b^{A}(u^{B}_{t_{2}}))|\mathcal{F}_{t_{1}}\big], \nonumber\\
           & & \{V^{l-1, B}(t_{1},\mathcal{Y}_{t_{1}},u^{B}_{t_{1}},b^{A}(u^{B}_{t_{1}}))- c_{t_{1}}^{Z}\}\bigg)
\end{eqnarray}
 where $SW^{i,Z}(.)$ is the so called \emph{intervention/switching operator} that quantifies the switching regime value and  it represents the obstacle in our RBSDE's formulation, while $l\in \{1,\dots,M\}$ denotes the number of switching time left.\\
 By running the program (29)-(30) backward over time one needs to keep trace - over each switching time - of both the players switching strategies - optimal or not - and the relative payoffs in order to calculate in $t=0$ the players value functions $J^{i}(.)$ and their game strategies (using definition 2.4.3). Then by checking the conditions (27)-(28),  the existence (and uniqueness) of the \emph{Nash equilibrium point} for the game (23) can be established.
  Definitely the equilibrium existence needs a careful numerical analysis and algorithm implementation that we leave for a future paper.

\subsection{\large Game solution in a special case and further analysis}

In this section we make some further reasoning on the game characteristics in order to possibly
simplify our general formulation (23) and to search for a solution.
In particular, we focus the analysis on the following three main points - already mentioned in
the past section - that have impact on the equilibrium characterization and existence:\\

\noindent a) \emph{information set between the players/counterparties};\\
b) \emph{rules of the game};\\
c)  \emph{differences in the objective functionals of the players/counterparties}.\\

\indent a) Firstly, a careful analysis of the game information set is fundamental to characterize and understand the game itself and  its equilibrium. In our game formulation (23) we have assumed symmetry in the information available for the players which helps to simplify the analysis, but in general one needs to specify what is the information available to them at all the stage of the game. Given that we have been working under the market filtration $(\mathcal{F}_{t})_{t\geq 0}$, under symmetry we get that both player knows $\forall\: t \in[0,T] $ the values of the market variables and processes that enter the valuation problem, namely
      \begin{equation*}
         \mathcal{F}_{t}^{A}  = \mathcal{F}_{t}^{B} =\mathcal{F}_{t}  \;\; \forall\: t \in[0,T]
      \end{equation*}
      So, both players are able to calculate the outcomes/payoff of the game through time. This implies that the players know each other cost functions so that the game is said \emph{information complete}\footnote{They know strategies and payoff at every stage of the game.} and it is easier to solve for a NEP knowing the \emph{best response function}.\\
      It is important to underline that the game is played simultaneously at the decision times but it is dynamic and recursive because of the optimal strategy played today will depend not only on the initial condition (that is usually \emph{common knowledge}) but on future decisions taken  by both the players. Clearly,  this complicates game characteristics of the game  imposing a backward induction procedure to search for an equilibrium point.\\
      Of course, the assumption to know the counterparty cost function is quite strong for our problem in which the parties of the underlying contract can operate  in completely different markets or industries, but it can be not uncommon to verify a \emph{cooperative behavior} between them. In particular, in \emph{cooperative games} the players aim to maximize or minimize the sum of the values of their payoff over times, namely \begin{eqnarray*}
                J^{coop}(y,u^{*}) &:=& \inf_{u^{*} \in \{\mathcal{C}_{ad}^{A} \cup \mathcal{C}_{ad}^{B} \}}\bigg[J^{A}(y,u^{A}) + J^{B}(y,u^{B})  \bigg]\\
                &:=& \inf_{u^{*} \in \{\mathcal{C}_{ad}^{A} \cup \mathcal{C}_{ad}^{B} \}} \mathbb{E}\bigg[\sum_{j} \int_{t}^{\tau_{j}^{i}\wedge\tau_{j}^{-i}\wedge T} B_{s} \Big[ F^{A}(y_{s},u^{A}, b^{B}(u^{A})) +\\
                &+&F^{B}(y_{s},u^{B}, b^{A}(u^{B})) \Big]ds + \big(l^{A}(\tau_{j}^{A} z_{j}^{A}) + l^{B}( \tau_{j}^{B} , z_{j}^{B}) \big)\bigg]\\
                &+& \big(G^{A}(y(T), b^{B}(u^{A})) + G^{B}(y(T), b^{A}(u^{B}))\big) \bigg| \mathcal{F}_{t} \Bigg]\;\; for \:j = 1,\dots,M.
            \end{eqnarray*}

      This type of equilibrium,  depending on the type of game considered, is much more difficult to study in the stochastic framework, given the necessity to find the condition under which players cooperate over time and have no incentives ''to cheat''  playing a different (non cooperative) strategy. This can be an interesting further generalization for our game model that would be important to examine in major depth. Hence this is an other interesting topic to study in relation to our type of game that would be important to examine this issue in major depth.\\

\indent b) Also the rules of the game are important in order to simplify the search for the equilibrium. In our model both the counterparties are able to switch optimally every time over the contract life. Discretizing the time domain, we have been lead to think at the game as played simultaneously through time over the switching time set that can be predefined in the contract or model specific. In terms of game theory, this means that a given decisional node of the game the players make their optimal choice based on the information available (which is \emph{common knowledge}) at that node and at the subsequent node they observe the outcome of the last interaction and update their strategy.\\
      An other possibility that may help to simplify things, is to assume that - by contract specifications - the counterparties can switch only at predefined times and that the two sets have null intersection, namely
      \begin{equation*}
       \big\{ \tau^{A}_{j}\cap \tau^{B}_{j}\big\}_{j=1}^{M} = \varnothing.
      \end{equation*}
      This happen for example if the right to switch is set as "sequential". So, again in terms of game theory, the strategic interaction and the game become \emph{sequential}: under the assumption of \emph{incomplete information}, this type of games are generally solved via backward induction procedure and one can search for a weaker type of Nash equilibrium.
      Clearly, we remind that in our case this type of equilibrium need to be
      studied under a stochastic framework which remains a cumbersome and tough task both
      analytically and numerically\\
      Clearly, we remind that this type of equilibrium should be studied under a stochastic framework which is analytically  more difficult and also numerically it can be a cumbersome task. A strategic sequential interaction like that, can be also obtained by setting time rules like the so called \emph{grace periods} within the  contract CSA, namely a time delta  $\Delta t$ that the other party has to wait - after a switching time -  before making its optimal switching decision.\\

\indent c) The last point really relevant in our game analysis  concerns the relative differences between counterparties objective functionals. In particular, recalling our model specifications, the main variables that have impact in this sense are:
      \begin{itemize}
        \item  differences in the default intensities processes $\lambda^{A}_ {t} $, $\lambda^{B}_{t} $;
        \item  differences in the cost function thresholds $ \delta^{A}(.)$, $\delta^{B}(.)$;
        \item  differences in funding/opportunity costs $R^{A}_{t}(.)$, $R^{B}_{t}(.)$;
        \item  differences in the (instantaneous) switching costs  $c^{z,A}_{t}(.)$, ,$c^{z,B}_{t}(.)$.
      \end{itemize}
      To be more clear, let us focus on some specific case.\\

      \textbf{1) Symmetric case}. Let us simplify things by considering the special case of our game (23)   in which symmetry between the party of the contract is assumed.
      In this special case,  we are able to show the existence of the solution for the game and we also highlight the impact on the equilibrium of just a simple constant threshold $\delta$ in the running cost functions. Under \emph{symmetry}, it is easy to show that the game
solution coincides with that of a stochastic control problem of switching type. In fact,  solving the control problem from just one player's perspective is equivalent to a game played by symmetric players with objective functional having the same parameters. In economic terms, the reason to consider a game between two \emph{symmetric} players can
be justified if one thinks to two institutions with similar \emph{business characteristics} other
than risk worthiness, that operate in the same country/region/market with the objective
to optimally manage the counterparty risk and the collateral and funding costs by signing
a contingent CSA in which are defined all the relevant parameters necessary to know each
other objective functional.\\
Hence, let us be more formal and let's consider a game played under these special symmetric conditions, it is not difficult to see that the game payoffs will be the same for both the player: in fact, being $\delta^{A}(.)=\delta^{B}(.)=0$, the square of BCVA and collateral costs functions is the same and also the instantaneous costs are assumed equal. This implies that also the best response functions will be equal for both the player, namely they play the same switching strategy, however the game is played simultaneously or sequentially. So, on the basis of this chain of thought, and imposing the following technical conditions\footnote{We refer here to section 2.1 of \cite{Dje08}.}
\begin{itemize}
  \item[\textbf{Hp1})] the stochastic factors that drive the dynamic of the system $(X_{t})_{0\leq t\leq T}$ and $(\lambda_{t})_{0\leq t\leq T}$, (that we indicate with the vector $\mathcal{Y}_{t}$ for brevity) are $   \mathbb{R}$-valued processes adapted to the market filtration $\mathbb{F}_{t}^{x,\lambda}= \sigma\{ W_{s}^{x,\lambda},s\leq t\}_{t\in[0,T]}$ assumed right-continuous and complete;
  \item[\textbf{Hp2})] the cost functions $F_{Z}^{i}(.) \in \mathcal{M}^{p}$ and $G_{Z}^{i}(.) \in \mathcal{M}^{p}$ while the switching costs $l^{i}_{Z}(.)^{Z}\in \mathcal{K}^{p}$, being deterministic and continuous;
  \item[\textbf{Hp3})] the running costs functions need to satisfy the \emph{linear growth condition} and for the switching costs $c_{Z}^{i}$ a technical condition as $\min\{c^{i}_{z}, c^{i}_{\zeta} \} \geq C $ for  $i\in\{A,B\} $ $t\leq T\wedge\tau$, switching indicators $\{z,\zeta\}\in Z$   and real constant $C>0$, is imposed in order to reduce the convenience to switch too many times;
\end{itemize}
the following result states.\\

\textbf{Proposition 2.6.1 (NEP Existence and uniqueness in symmetric case.)}. \emph{Assume
symmetric conditions for our Dynkin game of switching type (23), taking relations (24-26)  with the equality  and set  $\delta^{A}(.) =\delta^{B}(.)  = 0$. In addition, under the technical (Hp1 -Hp3)  exists and is unique a Nash equilibrium for this game and it coincides with the value function of the following stochastic control problem\footnote{The problem is explicitly set in \cite{Mot13}.}  that is
\begin{eqnarray}
          J(y,u) &=& \inf_{ u  \in\{\mathcal{C}_{ad}\}} \mathbb{E} \Bigg[\sum_{j} \int_{t}^{\tau_{j}\wedge T} B_{s}\big[F_{Z}(y_{s},u)\big]ds
          +   \sum_{j\geq 1} c_{z_{j}}(t)\mathbbm{1}_{\{ \tau_{j} <T\}} + G(y_{T}) \bigg| \mathcal{F}_{t} \Bigg]
        \end{eqnarray}
where the model dynamics $d\mathcal{Y}_{t}$ has been defined in section 2.3 and the following relation for the value function states
\begin{equation}
        J^{A}(y, u^{*}_{A}  ;u^{*}_{B} )  = J^{B}(y, u^{*}_{A}  ;u^{*}_{B} ) = V^{*}(y,u^{*}).\\
      \end{equation}
which indicates the irrelevance of the strategic interaction under symmetry.}\\

\emph{Proof.} The proof is easy given the above reasonings. In fact, under the \emph{symmetry conditions} and recalling the notation from the general game formulation (23) we have that the following relations hold:
      \begin{eqnarray*}
         F^{A}_{Z}(y_{s},u^{A}, b^{B}(u^{A}))&=& F^{B}_{Z}(y_{s},u^{B}, b^{A}(u^{B}))\\
         l^{A}_{z_{j}^{A}}(t)\mathbbm{1}_{\{ \tau_{j}^{A} \wedge \tau_{j}^{B} <T\}} &=& l^{B}_{z_{j}^{B}}(t)\mathbbm{1}_{\{ \tau_{j}^{B} \wedge \tau_{j}^{A} <T\}} \\
        G^{A}(y(T), b^{B}(u^{A})) &=& G^{B}(y(T), b^{A}(u^{B}))
      \end{eqnarray*}
        and being control sequence optimal for both players we can set $\tau_{j} := \tau^{A}_{j} = \tau^{B}_{j}$ which implies $\mathbbm{1}_{\{ \tau_{j}^{A} \wedge \tau_{j}^{B} <T\}} = \mathbbm{1}_{\{ \tau_{j} <T\}}$, namely $u^{*}_{A}= u^{*}_{B}$, which implies also the equality of the best response functions $b^{A}(u^{B}) =b^{B}(u^{A}) = 0$, given the assumptions on the thresholds $\delta^{i}(.)$. This means also that the strategic interaction becomes irrelevant and the game solution can be reduced to that of an optimal switching control problem equivalent for both the players, so that game (23) is reduced to the problem (31), namely
        \begin{eqnarray*}
          J(y,u) &:=& J^{A}(y,u^{A}, u^{B}) = J^{B}(y,u^{B},u^{A}) \Longrightarrow\\
          J(y,u) &=& \inf_{ u  \in\{\mathcal{C}_{ad}\}} \mathbb{E} \Bigg[\sum_{j} \int_{t}^{\tau_{j}\wedge T} B_{s}\big[F_{Z}(y_{s},u)\big]ds
          +   \sum_{j\geq 1} c_{z_{j}}(t)\mathbbm{1}_{\{ \tau_{j} <T\}} + G(y_{T}) \bigg| \mathcal{F}_{t} \Bigg].
        \end{eqnarray*}
By the proof of value function's existence and uniqueness  $V^{*}(y, u^{*})$ for this problem (for which we refer  to \cite{Dje08}), one derives the optimal sequence of switching times and indicators $u^{*}= \{\mathcal{T}^{*}, \mathcal{Z}^{*} \}$. that satisfies conditions (27) and (28) of NEP definition 2.5.1,
        being the control strategy optimal for both players (by symmetry). Indeed, given that  the two problems
    representation are actually the same, the NEP exists and is unique  - from the existence and
uniqueness of the value $V^{*}(y; u^{*})$ - and equation (32) is true, as we wanted to show. $\square$\\

      2) \textbf{Case $\delta^{A} =\delta^{B} \neq 0$}. In general with different function parametrization between players and incomplete or asymmetric information, the equilibrium is much harder to find and different strategies has to be checked. To give an idea of this, let us consider just a slight modification of the symmetric case conditions, setting for example the cost functions threshold $\delta^{A}= \delta^{B}>0$, and keeping the information incomplete and the game play simultaneous. By the symmetry of BCVA and (running) \emph{collateral/funding costs}, we know that a positive value of one term for $A$, is negative for $B$ and viceversa. So introducing the threshold create different payoffs for the player, as we can easily see below
      \begin{equation*}
        (BCVA^{A} - \delta)^{2} \gtrless (BCVA^{B} - \delta)^{2}
      \end{equation*}
      given that if $BCVA^{A}_{t}>0$ then $BCVA^{B}_{t}<0$ and viceversa\footnote{This is true also for the other switching costs related to collateral and funding. Of course one should consider also the weight of the expected payoff value from keeping the strategy for an other period.}. So, even though they knew each other objective functional, there would be some paths and periods in which the strategic behavior of the players is in contrast, say of \emph{war type} and others of \emph{peace type} (as by pure strategy definition 2.4.3), making the analysis more complicate.\\

     3) \textbf{Game banal solution case. } Worth of mention the possibility that the game is not played, namely it reveals to be never optimal to switch for both the players. It's relevant to study the conditions under this kind of behavior of the solution come up, given that the scheme would lose its \emph{economic sense}. This \emph{singular game solution} can come up if we formulate to our simultaneous game as a \emph{zero sum game}. This can happen by considering - for example - linear objective functionals with threshold $\delta\approx 0$, in fact - by symmetry of the BCVA and of funding cost function - a positive outcome for one player is negative for the counterpart\footnote{Obviously, when the value of the cost functions compensate each other, no switch take place}. Assuming instantaneous switching costs $c^{z}>0$ for both players and - to simplify - that both know  each other cost functions, we get that this game will never be played. The reason is that by the game zero sum structure, the optimal strategy for one is not optimal for the other, so every switch will be followed by the opposite switch at every switching time, like the sequence
     \begin{equation*}
        \big\{z_{1}=1,z_{2}=0,z_{3}=1,\dots,z_{M}=1\big\}
     \end{equation*}
     But by rationality and taking in account the positive cost of switching, one can conclude that the game will never be played by a rational agent.\\
     So, let us summarize this last logic chain of thoughts in the following proposition.\\

     \textbf{Proposition 2.6.2 (Game banal solution in the zero-sum case)}. \emph{Let us assume that game (23) be a zero-sum game with linear functionals set for both the counterparties. Assuming in addition the same funding costs for both players, $\delta \approx 0$ and positive switching costs $c^{z}>0$ (for both), then the optimal strategy is to never play this game (that is a banal solution of the game).}\\

We end the section by remarking the importance of the points highlighted in the construction
of some kind of equilibrium for a Dynkin game of switching type as our one. Although mainly
theoretical, the existence of the equilibrium and the derivation of the conditions under which a
non banal solution exists, are relevant economically and in the contract design phase.\\
Hence, the main task to pursue in future research are a rigorous proof of the existence and of
the equilibrium for this type of game, and the definition of an efficient algorithm to check the
model solutions.\\

\section{\Large Applications and further researches}
Switching type mechanisms like the one we have analyzed can find different applications into the
wild world of finance. The basic underlying idea is to ensure
\emph{flexibility} to agents' investments decisions over time which is a usual objective in \emph{real option theory}. Our problem has been thought
mainly in a risk management view but with the development of new techniques and algorithms,
also the related pricing problem will be tackled efficiently and more financial contract would find
useful and convenient - in a optimal risk management view - this type of contingent mechanisms.
As regards just a possible further application in risk management, it would be important to
deepen the analysis of a switching type collateralization from a \emph{portfolio perspective}, taking for
example the view of a \emph{central clearing}. In particular, it would be relevant to show possibly analytically but mainly with numerical examples, the greater convenience of the switching/contingent
solution respect to a\emph{ non contingent/standard collateral} agreement like the \emph{partial or full} one,
including clauses like, \emph{early termination}, \emph{netting} and others. This is an hard program, which
needs a generalized model formulation in order to include all the CSA clauses and in order to
deal with the high recursion that characterizes the problem.\\
In a pricing view, we remind the example - of the   fixed income market - some particular bonds
called \emph{flippable} or \emph{switchable}, that are characterized by options to switch the coupon from fix to
floating rate. Clearly, in this case the valuation is easier given that this securities have a market
and are not traded OTC so one does not need to include in the picture counterparty risk, funding and CSA cash
ows. Anyway, it would be interesting to delve into the valuation of an OTC
contract in witch also the dividend flows can be subject to contingent switching mechanism. As
regards similar case, a problem that can be very interesting and difficult to tackle is the valuation of a
flexi swap in presence of a contingent collateralization like our one.\\
The main characteristics of a flexi-swap are:\\
 a) the notional of the flexible swap at period $n$ must lie (inclusively) between predefined bounds $L_{n}$ and $U_{n}$;\\
  b) the notional of the flexible swap at period n must be less than or equal
to the notional at the previous period $n-1$;\\
  c) The party paying fixed has the option at the start of each period $n$ to
choose the notional, subject to the two conditions above.\\
In other words, we deal with a swap with  multiple embedded option that allows one party to change the notional under certain constraints defined in the contract. This kind of interest rate swaps are usually used as hedging instruments of other  swaps having notional linked to loans, especially mortgages.\footnote{ In this sense is like there were a third reference represented by the pool of loans.} The underlying idea is that the fixed-rate payer (the option holder) will amortize as much as allowed if interest rates are very low, and will amortize as little as allowed if interest rates are very high.\\
Given a payment term structure  $\{T_{n}\}^{N}_{n=0}$ and a set of coupons $X_{n}$ (with unit notional) fixing in $T_{n}$ e paying in $T_{n+1}$ ($n=0,1,\dots,N-1$), the \emph{flexi swap} is a fixed vs floating swap where the fixed payer has to pay a net coupon $X_{n}R_{n}$ in $T_{n+1}$,  the notional $ R_{0}$  is fixed upfront  at inception and for every $T_{n}$,  $R_{n}$ can be amortized if it respects some given constraints defined as follows:
\begin{enumerate}
  \item  deterministic constraints : $R_{n} \in [g_{n}^{low}, g_{n}^{high}]$;
  \item local constraints function of the current notional: $R_{n} \in [l_{n}^{low}(R_{n-1}) l_{n}^{high}(R_{n-1}) ]$;
  \item market constraints (libor, swap denoted with $X_{n}$): $R_{n} \in [m_{n}^{low}(X_{n}), m_{n}^{high}(X_{n}) ]$.
\end{enumerate}
The valuation procedure of this type of swap involves a backward recursion keeping track of the notional in every payment date.  But introducing also the switching collateralization, the valuation become an "\emph{intricate puzzle}" given the recursive relation between the optimal switching strategy, the price process of the claim which in addition depends on the optimal  notional choice over time.  Simplifying modeling assumptions are needed to break the curse of recursion in a defaultable OTC contract like this.\\\\

\section{\Large Conclusions}
In this work, we have generalized the contingent CSA scheme defined in our preceding work to
the bilateral case allowing the strategic interaction between the counterparties of a defaultable
(OTC) contract. The problem in this case has a natural formulation as a stochastic differential
game - a generalized Dynkin game - of switching type, for which - as far as we know - no analytical
solution for a \emph{Nash equilibrium point} is known.\\
We have shown, in particular, that the game solution is strictly related to that of a system of
reflected BSDE with interconnected barriers and generator functions. Only by imposing strong assumptions and simplifications we are able to prove the game solution, in the so called \emph{symmetric case}.
Further research  are needed and addressed in the end in the field of stochastic games and RBSDE and some
interesting applications in finance are highlighted in order to show also the importance in practice
of our mainly theoretical problem.

\end{document}